\documentstyle[11pt]{article}
\raggedright
\def\bbbr{{\rm I\!R}}
\def\bbbc{{\mathchoice {\setbox0=\hbox{$\displaystyle
\rm C$}\hbox{\hbox
to0pt{\kern0.4\wd0\vrule height0.9\ht0\hss}\box0}}
{\setbox0=\hbox{$\textstyle\rm C$}\hbox{\hbox
to0pt{\kern0.4\wd0\vrule height0.9\ht0\hss}\box0}}
{\setbox0=\hbox{$\scriptstyle\rm C$}\hbox{\hbox
to0pt{\kern0.4\wd0\vrule height0.9\ht0\hss}\box0}}
{\setbox0=\hbox{$\scriptscriptstyle\rm C$}\hbox{\hbox
to0pt{\kern0.4\wd0\vrule height0.9\ht0\hss}\box0}}}}
\def\bbbone{{\mathchoice {\rm 1\mskip-4mu l} {\rm 1\mskip-4mu l}
{\rm 1\mskip-4.5mu l} {\rm 1\mskip-5mu l}}}
\renewcommand{\Re}{\rm Re\,}

\newcommand{\cB}{{\cal B}}

\newcommand{\cH}{{\cal H}}
\newcommand{\cT}{{\cal T}}

\newcommand{\cW}{{\cal W}}
\newcommand{\rL}{{\rm L}}
\newcommand{\rR}{{\rm R}}
\newcommand{\para}{{\scriptscriptstyle \,\|}}
\newcommand{\ba}{{\bf a}}
\begin{document}
\title{Connections and Metrics \\Respecting Standard Purification}
\author{J.~Dittmann$^*$\and
A.~Uhlmann$^{**}$}
\date{Universit\"at Leipzig, Germany\\
$ $\\$^*$Mathematisches Institut,
$^{**}$Institut f.~Theoretische Physik\\
$ $\\$ $\\
June 4, 1998}
\maketitle
\begin{abstract}
\noindent
Standard purification interlaces Hermitian and Riemannian metrics on the space
of density operators with metrics and connections on the purifying
Hilbert-Schmidt space.
We discuss connections and metrics which are well adopted to purification, and
present a selected set of relations between them. A connection, as well as a
metric on state space, can be obtained from a metric on the purification space.
We include a condition, with which this correspondence becomes one-to-one.
Our methods are borrowed from elementary $^*$-representation and fibre space
theory.
We lift, as an example, solutions of a von Neumann equation, write down
holonomy invariants for cyclic ones, and ``add noise'' to a curve of pure
states.
\end{abstract}
\section{Introduction}
In \cite{Pe96}, see also \cite{PS96}, the monotone
Hermitian and Riemannian metrics in the (finite dimensional)
spaces of all density operators are classified. Based on the
theory of operator means, \cite{KA80}, they are indexed
by a real function, $f$, operator monotone
on $(0 , \infty)$. These metrics play an important role in
domains like quantum information geometry,
quantum versions of statistical estimation and decision
rules,  \cite{book82.5}, \cite{IJKK82}, \cite{OP93}.

D.~Petz communicated his main results
to us prior to publication, and about that time
we started to ask for the effect of a purifying lift
to these metrics. There are clear reasons for this.
One of the present authors,
(A.U.), had defined 1986 in \cite{Uh86a} an extension of the geometric
phase, \cite{Be84}, \cite{Si83}, see also \cite{AA87}, \cite{AS87},
to curves of density operators by the help of a ``parallelity
condition''.
The condition singles out, up to a global gauge
(or a global partial isometry),
a distinguished ``parallel lift'' within all purifying lifts of a curve of
density operators.
It turns out, \cite{Uh91a}, that a connection form
(a gauge potential), here called $\ba^{\rm geo}$,
is governing the transport of the purifying vectors, such that the
parallelity condition results from the request for horizontality.
In 1992 G.~Rudolph and one of the
authors, (J.D.), considered a large class of gauge potentials,
including $\ba^{\rm geo}$, which
rests on a purification scheme and which enables variants of the
geometric phase along curves of density operators.
It seems natural to ask for a link between these objects:
(a) the connection forms just mentioned, (b) certain
Hermitian (Riemannian) metrics on the purification space, and,
if respecting the symmetry of the scheme, (c) metrics induced
from (b) on the space of density operators.

Purification is essentially representation theory of observables
and of the algebra in which they are contained. Principally one may
use any unital $^*$-representation of
the ``algebra of observables'' over which
the states can be defined. Its Hilbert representation space should
only be large enough to allow for a representation of the states by vectors.
If this condition is fulfilled,
transport mechanism, its non-commutative phases, metrics, and other
geometric objects can be constructed by relying on their form and
appearance in the pure state case.

In our paper we remain within an elementary setting:
Our density operators live on an
Hilbert space ${\cH}$ of finite dimension $n$. In our
convention, a density operator should not necessarily be normalized.
We speak of ``density operators'' whether their trace is one or not.
The algebra of observables is the algebra $\cB(\cH)$ of all operators
acting on $\cH$. The representation or purification space,
$\cW$, is identified with the algebra of operators and equipped
with the Hilbert-Schmidt scalar product. (In infinite dimensions
$\cW$ will be the space of Hilbert-Schmidt operators.) We try to
emphasize the different meaning of operators by different notations:
Operators acting on $\cH$ are denoted by small, those acting on
$\cW$ often by capital letters. (Some authors call the operators of
$\cB(\cW)$ ``superoperators''.) The next section is devoted to explain our
notation in more details. In our paper purification takes
place in the standard representation of $\cB(\cH)$, i.~e.
in the GNS-representation based on the trace. For that reason we called
it {\it standard purification}. In section 3 the formalism is
extended to velocity vectors, i.~e.~to tangents, at density operators
and at their purifications. Purification defines vertical
tangents in a canonical way. A tangent, orthogonal to the space of
vertical tangents, is called horizontal, provided the tangent spaces
carry a real Hilbert space structure, i.e.~a Riemannian metric.
Equivalently, within all purifying lifts of a given curve of density
operators, those with the least length are horizontal.

Section 4 exemplifies our task in defining horizontality by the
real part of the Hilbert-Schmidt metric. As one knows, the Bures
length of a curve of density operators and the Hilbert-Schmidt
length of an horizontal lift are equal one to another.
In deriving the parallelity condition we meet
 some peculiarities with tangents of purifying vectors if they belong
to density operators with some vanishing eigenvalues. The reader
will find a short account of the
relation between the connection form $\ba^{\rm geo}$, \cite{Uh91a},
governing the geometric phase, and the Riemannian Bures metric.

Indeed, it last some time to ask and to give an affirmative answer
to the question, whether the topological metric of Bures is Riemannian
\cite{Uh92b}, \cite{Hu92b}, \cite{Uh92a}.  Essential differential
geometric properties are in \cite{Di93a}, see also \cite{Hu93a}
for $\dim \cH = 3$. Relations to quantum information theory can
be seen in \cite{BC94}, \cite{FC95}.
However, a parameterization in terms of the operators' matrix
elements remains cumbersome, except $\dim \cH = 2$.

Concerning $\ba^{\rm geo}$, which extends the geometric phase
to (closed) curves of density operators, an example is in
the last section. There is a further issue, to be mentioned
at least: The gauge potential for the 2-dimensional density
operators, \cite{DR92b}, living on a 4-dimensional purification
space, satisfies the Yang-Mills equations.
With a certain cosmological
constant, it even is a solution of the combined
Yang-Mills-Einstein equations \cite{To93}. Meanwhile we know,
\cite{Di98}, $\ba^{\rm geo}$ satisfies the Yang-Mills equations
for every finite dimension of the supporting Hilbert space $\cH$.
These findings
may be seen as extensions to mixed states of numerous examples
relating the original Berry phase to Dirac monopoles, and
the Wilczek and Zee phase, \cite{WZ84}, to instantons.

Section 6 is devoted to the class of connections introduced
in \cite{DR92a}, which are, so to say, ``relatives'' of $\ba^{geo}$,
compatible with the purification scheme.
They are characterized by a function
$F$, defined on $(0,\infty)$, and fulfilling
$\bar F(1/t) = -F(t)$. Some equations become more appealing
by using the function $r$, the arithmetic mean of $F$ and $1$.
The  connections forms $\ba$ assign to every tangent $x$
at the lift $w \in \cW$ of $\varrho = ww^*$
a value in the Lie algebra of U$(n)$.
The action of the gauge group induces the ``canonical''
connection $\ba^{can}$. The canonical connection
is gained with the choice $F = 0$. The connection $\ba^{geo}$
is constructed with $F(t) = (t-1)/(t+1)$. As we shall see, only
a connections with real $F$ can be obtained from an appropriate
Hermitian metric. We believe, the complete class is a more natural
object at the complexified tangents. They all
decompose as $\theta - \theta^* $ with $\theta$ of type (1,0).

We specify the class of Hermitian metrics by another positive and
real valued function, $k$, on the positive half-axis. The metrical
form for the tangents at a purifying vector, $w$, will be given by the
inverse of the (``super'')operator $k(\Delta_w)$, where
$\Delta$ is the field of modular operators.
There is an antilinear
operator, a modification of Tomita-Takasaki's $S_w$-operator, which
admits just the horizontal tangents as fix points. The connection
adjusted to the metric is characterized by various relations between
the functions $k$, $F$, and $r$. Moreover, every one of the
Hermitian metrics considered on the tangent space of $\cW$ is a
lift of exactly one Hermitian form on the space of density operators.
The latter depends on a function $f$ which is related to $k$.
The Riemannian metric on the density operators is gained as the
real part of the Hermitian one, and it corresponds to the harmonic
mean of $f(t)$ and $tf(1/t)$.
 Further we discuss  an additional condition, which enables us to
assign  a unique connection form to a given  monotone Riemannian
state space metric.
These metrics are induced from the Hilbert-Schmidt metric by some
constraints
on the purifying vectors replacing the orthogonality condition of
the Bures case.

The starting point has been a set of connections, compatible with
the purification procedure, to define reasonable parallel
transports along curves of density operators. We return to this
issue in purifying horizontally solutions of von Neumann equations.
Cyclic solutions give rise to some holonomy invariants.
There are constraints on $F$ for extending the parallelity conditions
to the boundary, in particular to pure states. If they are fulfilled,
the holonomy invariants reduce to the well known geometric phase
of Berry for pure states. At the end we ask what happened
if ``noise'' is added to a closed path of pure states.

\section{ Standard Purification }
We start by reviewing some basic ideas of the purification
procedure.

Let $\cH$ be a complex Hilbert space of finite dimension $n$.
Following the usage in Physics we call $\langle .,. \rangle$
its scalar product and assume antilinearity in its left,
linearity in its right argument.

$\cB(\cH)$ denotes the $^*$-algebra of linear operators acting
on $\cH$. A {\it state} is a positive linear form
over the algebra which takes the value 1 at the identity
of $\cB(\cH)$. Generally, a linear form $l$ over our
algebra is uniquely represented by

\begin{equation} \label{linform}
l(b) = {\rm Tr} \, b \omega, \quad \forall \, b \in \cB(\cH)\,.
\end{equation}

The linear form is positive if and only if $\omega$ is a positive
element of $\cB(\cH)$. We then call $\omega$ a {\it density operator}
to come in accordance with its usage in physics. A density
operator represents a state iff its trace is one.

A {\it purification} of a positive linear form over $\cB(\cH)$
is a lift to a pure linear form of a larger algebra.

A way, to do so, is that: With another, auxiliary Hilbert space
$\cH^{\rm aux}$, with at least the same dimension, we consider

\begin{equation} \label{puri01}
\cH  \otimes \cH^{\rm aux}, \quad n = \dim \cH \leq \cH^{\rm aux}
\end{equation}

and the inclusion (which, indeed, is a $^*$-representation,)

\begin{equation} \label{puri02}
\cB(\cH) \hookrightarrow \cB(\cH)
\otimes 1^{\rm aux}
\end{equation}

into the operator algebra of the Hilbert space (\ref{puri01}).
Let $\varrho$ be the density operator of a positive linear
form $l$ over $\cB(\cH)$. A vector $\psi$ of (\ref{puri01})
is said to {\it purify} $l$, and hence $\varrho$, iff

\begin{equation} \label{puri04}
l(b) \equiv {\rm Tr} \, b \varrho
= \langle \psi, b \otimes 1^{\rm aux} \, \psi \rangle
\quad \forall \, b \in \cB(\cH)\,.
\end{equation}

A distinguished way to choose the auxiliary Hilbert space is to
require

\begin{equation} \label{spu01}
\cH^{\rm aux} = \cH^*, \quad \cW := \cH \otimes \cH^*\,,
\end{equation}

which results in the {\it standard purification}, based on the
standard representation of $\cB(\cH)$. In what follows this
choice is assumed, and we have to fix some notations and
conventions at the beginning.

Let $\phi \in \cH$. The element $\phi^* \in \cH^*$,
is defined by $\phi^*(\phi') = \langle \phi, \phi' \rangle$.
In Dirac's notation:

\begin{equation} \label{spu02}
\phi \leftrightarrow |\phi\rangle, \quad
\phi^* \leftrightarrow \langle \phi|\,.
\end{equation}

Being in finite dimensions, every operator is Hilbert-Schmidt, and
$\cW$ is canonically isomorphic to $\cB(\cH)$. This can be made
explicit with two arbitrarily chosen orthonormal bases
$\phi_1, \phi_2, \dots$ and $\phi'_1, \phi'_2, \dots$ of $\cH$
in writing

\begin{equation} \label{spu03}
w = \sum
|\phi_j \rangle \langle \phi_j, w \, \phi'_k \rangle \langle \phi'_k|,
\quad w \in \cW\,.
\end{equation}

The Hilbert Schmidt scalar product on $\cW$ is

\begin{equation} \label{spu04}
(w_2, w_1) := {\rm Tr} \, w_2^* w_1 =
\sum \langle w_2 \phi'_k, \phi_j \rangle \langle \phi_j, w_1 \phi'_k
\rangle\,.
\end{equation}

The star operation in $\cB(\cH)$ is equivalent with a conjugation
in $\cW$,

\begin{equation} \label{spu05}
w \to w^* \quad {\rm or} \quad (\phi \otimes \tilde \phi^*)^* =
\tilde \phi \otimes \phi^*\,.
\end{equation}

We need some operators acting on $\cW$. The standard representation
of $\cB(\cH)$ is the inclusion (\ref{puri04}), specified by
(\ref{puri02}), and acting as follows:

\begin{equation} \label{spu06}
b \mapsto L_b, \quad L_b w := b w, \quad b \in \cB(\cH)\,.
\end{equation}

We also need the right multiplication $R_b$, i.e. $R_b w = wb$.
The right multiplication can be used to
implement the standard representation of $\cB(\cH^*)$. Notice the
different meaning of the $^*$-operations on $\cW = \cB(\cH)$ and
on $\cB(\cW)$ seen in

$$
(L_b)^* = L_{b^*}, \quad (L_b w)^* = (R_b)^* w^*
$$

and in similar relations after exchanging $L_b$ and $R_b$.
Now, let $\hat l$ be a linear form on $\cB(\cW)$ and
$l$ its {\it restriction} or {\rm reduction} onto $\cB(\cH)$.
The relation

\begin{equation} \label{spu07}
\hat l \mapsto l, \quad l(b) := \hat l(L_b), \quad b \in \cB(\cH)
\end{equation}

encodes the partial trace over $\cH^*$ on $\cW$.
Focusing our attention to the purification procedure,
we shall apply this well known mapping mainly to
linear functionals of rank one.
In that case the essence of the reduction mapping to the factors of
$\cW$ is contained in

\begin{equation} \label{spu09}
(w_2, L_b R_c w_1) = {\rm Tr} \, w_2^* b w_1 c\,.
\end{equation}

Its left-hand-side defines a linear form $B \mapsto (w_2, B w_1)$
over $\cB(\cW)$, and, varying $w_1$ and $w_2$ within $\cW$,
one can get every linear functional of rank one.
Presently we need to consider (\ref{spu09}) with
$w_1=w_2=w$ and with either $c$ or $b$ the identity operator.
Then, for $B \in \cB(\cW)$ and $b, c \in \cB(\cH)$,
the left and the right side of (\ref{spu09}) may be rewritten
\begin{equation} \label{spu10}
\hat l(B) = (w, B w), \quad l(b) = {\rm Tr} \, w w^* b,
\quad l'(c) = {\rm Tr} \, w^* w c\,.
\end{equation}
$\varrho = \varrho_{l} :=ww^*$ is called the {\it density} or the
{\it density operator} of $l$, while $w$ is said to
{\it purify} $l$. In the same spirit, a positive linear functional
$\hat l$ of rank one, which reduces to $l$, is a {\it purification}
of $l$.

From now on, instead of switching forth and back between linear forms
and their densities, we remain mainly with the latter. Accordingly
we define the mappings

\begin{equation} \label{spu08}
\Pi \,  w = w w^*, \quad \Pi' \,  w = w^* w\,.
\end{equation}

The mapping $\Pi$ (and similarly the mapping $\Pi'$),
is slightly more subtle than the reduction mapping (\ref{spu07}).
Its domain of definition is $\cW$. Thus $\Pi$ is composed of a
Hopf bifurcation from $w$ to the rank one density operator $|w)(w|$,
representing the linear form $B \to (w, B w)$, followed by the
reduction (\ref{spu07}):

$$
w \longmapsto |w)(w| \longmapsto w \, w^*\,.
$$

Here we used Dirac's notation relative to the scalar product
(\ref{spu04}) in $\cW$.
$\Pi$ is a bundle projection, where the bundle
space is $\cW$ and the base space is the cone of (not necessarily
normalized) density operators (i.~e. positive trace class operators).
Being in finite dimension, the base space is the positive cone of
$\cB(\cH)$.
The bundle fibers are manifolds.
However, the dimension of the fibers vary with the rank $n_w$
of $w \in \cH$.
Therefore certain discontinuities occur if the rank is changing.

All this can be seen by the ``diagonal'' form of
(\ref{spu03}), which is the Gram-Schmidt decomposition of $w$.
Let $\lambda_1, \lambda_2, \dots$ be the $n_w$ non-zero eigenvalues
of $ww^*$ and $\phi_1, \phi_2, \dots$ their orthonormal
eigenvectors,

\begin{equation} \label{deco1}
w \, w^* = \sum \lambda_j |\phi_j \rangle \langle \phi_j|,
\quad \lambda_k > 0
\end{equation}

There exists exactly one other orthonormal basis of vectors,
$\phi'_1, \phi'_2, \dots$ of the same length $n_w$, fulfilling

\begin{equation} \label{deco2}
w = \sum \sqrt{\lambda_k} |\phi_k \rangle \langle \phi'_k|,
\quad w^* w = \sum \lambda_j |\phi'_j \rangle \langle \phi'_j|
\end{equation}

and the positive numbers $\lambda_j$ sum up to $(w,w)$.
From (\ref{deco2}) one can read off the polar decompositions

\begin{equation} \label{deco3}
w = \sqrt{ww^*} v = v \sqrt{w^*w},
\quad v = \sum | \phi_k \rangle \langle \phi'_k |\,.
\end{equation}

The index $k$ runs from 1 to $n_w$. One may call $v$ the
{\it phase of $w$ relative to} $\varrho = ww^*$.
The projection operators $v^*v$ and $vv^*$, attached to the
partial isometry $v$, map $\cH$ onto the support spaces of $w^*w$
and $ww^*$ respectively. Later on we need the
operator $J = J_w$,
\begin{equation} \label{modcon1}
J_w x = v x^* v =
\sum |\phi_j \rangle\langle \phi'_j, x^* \, \phi_k \rangle
\langle \phi'_k|\,,
\end{equation}
which, for completely entangled $w$, is the well known
{\it modular conjugation}. One easily establishes
\begin{equation} \label{modcon2}
(J_w)^2 x = (vv^*) x (v^*v), \quad (J x, y) = (J y, x)\,.
\end{equation}

If $\varrho > 0$ is a density operator, the set
$\Pi^{-1} \varrho$ consists of all $w$ satisfying $\varrho = w w^*$.
Along this fiber the orthoframe $\phi'_1, \phi'_2, \dots$
in (\ref{deco2}) and (\ref{deco3}) varies arbitrarily.
Thus the fiber at $\varrho$ is isomorphic, though
not canonically, to a complex Stiefel manifold. These isomorphisms
are parameterized by the different possibilities to choose an
orthoframe for the non-zero eigenvalues of $\varrho$.
The {\it structure}
or {\it gauge group} of $\Pi^{-1} \varrho$ consists of all
unitary $u \in \cB(\cH)$ acting by $R_u$.

Iff $\varrho$ is already pure, $\varrho = |\phi\rangle\langle\phi|$,
its purifications reads $w = |\phi\rangle\langle\phi'|$.
That is, the purifying vectors are necessarily product vectors
(``unentangled'' vectors).

In case the rank of $\varrho$ is larger than one, $w$
is called {\it entangled} in the domain of quantum
information theory. Accordingly, {\it complete entanglement}
of $w$ is reached if the density operator $\varrho$ is of maximal
rank $n_w = \dim \cH$. In this case, in traditional $^*$-representation
theory, $\varrho$ is called {\it faithful} and $w$ {\it separating}.
$\varrho = ww^*$ is faithful iff $w$ is invertible.

The set of all faithful $\varrho$ is the base space of a
principal fiber bundle with free action of the unitaries $R_u$.
The fiber space consists of all invertible $w$, the projection is
$\Pi$.

\section{Purification and Tangents}

A smooth, oriented curve in $\cW$, passing through $w$, defines
at $w$ a {\it tangent} or {\it velocity vector} $x$.
Hence the tangent space, $\cT_w$ at $w$, may be
identified with $\cW$ if considered as a real linear space.

Assume that $w$ and the unitaries $u$ depend smoothly on a
parameter, and let us use a dot to show parameter differentiation.
The gauge transformation $w \to w':= wu$ induces the relation

\begin{equation} \label{g01}
x \mapsto x' = x u + w \dot u, \quad x = \dot w, \, \,
x' = \dot w'\,.
\end{equation}

Let us now consider $\Pi$, and assume $\Pi w = \varrho$.
$\Pi$ induces a mapping $\Pi_*$ from the tangent space of $\cW$
into the density operator's tangents.

Being a first order problem, it is
sufficient for the following to assume a curve as simple
as possible, say $w(\lambda) = w + \lambda x  $. The curve is
projected by $\Pi$ to a curve of density operators
$\varrho_{\lambda} = w(\lambda) w^*(\lambda)$ of $\cB(\cH)$.
Differentiating at $\lambda = 0$ results in a tangent $\Pi_* x =\xi$
at $\varrho$.

\begin{equation} \label{tn01}
\xi = \dot \varrho, \quad \xi = (ww^*)^{\cdot} = x \, w^* + w \, x^*\,.
\end{equation}

A tangent vector $x$ at $w$ is called {\it vertical}
iff $\Pi_* x = 0$.
The real vector space of the vertical tangents at $w$
is denoted by $\cT_w^{\rm ver}$. It is a straightforward and well
known exercise to show: The gauge transformation $x \to x'$ of
(\ref{g01}) maps vertical tangents at $w$ to vertical tangents
at $w'$.

We look at vertical tangents as labels for
the physical phase. The phase of a single state or of its
density operator is not an observable.
Which purifying vector $w$
we choose, is physically irrelevant. What can be observed are
relative phases, for example in interference experiments.
The relative phases should
depend on the way a density operator is changed to become another one.
There should be a protocol according to which the tangents,
and hence the phases, are transported along a curve within the
space of density operators.
This can be achieved by the help of a parallel transport.

The standard procedure is to split the tangent space at every
$w$ into a direct sum of the vertical and of an horizontal part.
Respecting the complex linear structures,
we restrict ourselves to decompositions defined by
the real part of an Hermitian metric: We assume at every
$w$ a distinguished positive Hermitian sesquilinear form

\begin{equation} \label{tn03}
w \mapsto (x_2, x_1)_w, \quad x_1, x_2 \in \cT_w\,.
\end{equation}

For completely entangled $w$ it should be positive definite.
Now $\Re (.,.)_w$, the real part of (\ref{tn03}),
converts the tangent space at $w$ into a {\it real} Hilbert space.
The {\it velocity} with which a curve goes through $w$ is the square
root of $(x,x)_w$ with $x$ the tangent at that point. In this setting,
parallel transport is asking for a minimal velocity lift
of a given tangent at the base space. This, in turn, induces a
metrical structure at the base space: One calls {\it velocity of a
base space tangent} the minimum of the velocities of all possible lifts.

Thus, the {\it horizontal part}, $x^{\rm hor}$, of a tangent
$x$ at $w$ is the unique element of the set
$x + \cT_w^{\rm ver}$ with the smallest velocity.
This is in accordance with the definition of $\cT_w^{\rm hor}$
as the orthogonal complement of $\cT_w^{\rm ver}$ in the
real Hilbert space $\cT_w$, the latter equipped with the
scalar product $\Re (.,.)_w$.

There is a distinguished real subspace, $\cT_w^{\rm Ver}$,
within $\cT_w^{\rm ver}$ containing all tangents

\begin{equation} \label{tn02}
x = w a, \quad a = - a^* \in \cW\,,
\end{equation}

which are obviously vertical.

If $w$ is invertible (completely entangled),
every vertical tangent can be uniquely expressed in that way.
But generally, $\cT_w^{\rm Ver}$ is a proper subspace of
$\cT_w^{\rm ver}$. We call a vertical tangent {\it neutral} iff
it is orthogonal to $\cT_w^{\rm Ver}$ with respect to $\Re (.,.)_w$.
Hence, every tangent $x$ allows for an orthogonal decomposition

\begin{equation} \label{tnz}
x = x^{\rm hor} + x^{\rm ver}, \quad
x^{\rm ver} = x^{\rm neutral} + x^{\rm Ver}\,.
\end{equation}

\section{Phase transport and Bures Metric}

The most natural and simple choice for the Hermitian metric
$(x_2, x_1)_w$ of (\ref{tn03}) is certainly the Hilbert-Schmidt
scalar product (\ref{spu04}). This choice is particularly interesting
for several reasons.

At first it gives a straightforward generalization of the geometric
phase by the parallel transport evolving from this choice.
Indeed, one obtains a natural extension of the Fock \cite{Fo28},
Berry \cite{Be84}, Simon \cite{Si83}, Wilczek and Zee \cite{WZ84}
parallel transport to density operators.

Transport of state vectors along closed curves generates
an holonomy problem. In the period
between V.~Fock and M.~Berry this has become clear.
B.~Simons explained how to calculate the holonomy by the
second Chern class of the Hilbert space if considered as a line bundle.
There is an extensive literature on the transport of phases along
curves and loops of pure states, see \cite{books1} for a selection of
important results, applications, and references.
Particular examples in using and calculating the geometric phase
can be found already in papers decades past.

Secondly, one gets a Riemannian metric, \cite{Uh92b},
on the (not necessarily normalized)
density operators of $\cB(\cH)$. Its distance function
is the distance introduced by Bures \cite{Bu69} in following a similar
construction of Kakutani \cite{Ka48} in probability spaces.
Being the infinitesimal version of {\it Bures' distance},
we call this Riemannian metric {\it Bures metric}.

And, finally, already the choice
\begin{equation} \label{bu01}
(x_2,x_1)_w = (x_2,x_1), \quad \forall \, w
\end{equation}
shows essential problems in deviating from a genuine fiber bundle.

We start by enumerating the tangents $y$ orthogonal to
$\cT_w^{\rm Ver}$
$$
(y, wa) + (wa, y) = 0 \quad \forall \, \, a + a^* = 0\,.
$$
That condition straightforwardly comes down to

\begin{equation} \label{bu02}
y^*w = w^*y
\end{equation}

and $y$ is orthogonal to all Ver-tangents iff $w^*y$ is Hermitian.
(\ref{bu02}) is the {\it parallelity condition} \cite{Uh86a},
which extends the transport condition for the geometric phase
from pure to mixed states.

To decompose $y$ in its neutral and horizontal part, we start by
completing the two orthonormal systems of the Schmidt decomposition
(\ref{deco2}) arbitrarily and set $\lambda_j = 0$ if $j > n_w$.
By sandwiching (\ref{bu02}) between the orthobase $\{ \phi_i \}$
we get
$$
\sqrt{\lambda_k} \langle \phi_j, y^* \phi'_k \rangle =
\sqrt{\lambda_j} \langle \phi'_j, y \phi_k \rangle\,.
$$
There evolve two conditions on the matrix elements:
$$
j \leq n_w, \, \, k > n_w \, \, \Rightarrow  \, \,
\langle \phi'_j, y \phi_k \rangle = 0\,.
$$

$$
 k, j \leq n_w \, \, \Rightarrow  \, \,
{ \langle \phi_j, y^* \phi'_k \rangle \over \sqrt{\lambda_j} } =
{ \langle \phi'_j, y \phi_k \rangle \over \sqrt{\lambda_k} }\,.
$$
No restriction occurs for $j > n_w$, $k \leq n_w$.
There is an Hermitian $g$ such that

\begin{equation} \label{bu03}
\langle \phi_j, g \phi_k \rangle =
{ \langle \phi'_j, y \phi_k \rangle \over \sqrt{\lambda_k} },
\quad k \leq n_w\,.
\end{equation}

One may choose the matrix elements of $g$ with indices both larger
than $n_w$ arbitrarily but consistent with $g = g^*$.

{\sl The tangent $y_1 =gw$ is horizontal}, \cite{DG90}, \cite{Uh89b},
because it is orthogonal to all ver-tangents $x$. Indeed, $xw^*+wx^*=0$
implies $(gw,x)+(x,gw)=(g,xw^*+wx^*)=0$.  What remains
to check is the case of  a tangent $y_0$,  real orthogonal to all
$gw$, $g=g^*$, and to all Ver-tangents. From the first condition it
follows $wy_0^* + y_0w^* = 0$, hence verticality, and from the second
we obtain $w^*y_0 = y_0^*w$. This is equivalent with
$$
\langle \phi_j, y_0 \phi'_k \rangle = 0 \quad \forall \, j, k \leq n_w
$$
or

\begin{equation} \label{bu04}
y \, \, {\rm neutral} \, \Leftrightarrow \, w^* y = y w^* = 0\,.
\end{equation}

We conclude that every tangent $x$ allows for
a unique decomposition

\begin{equation} \label{bu05}
x = g w + x_0 + w a
\end{equation}

in an horizontal, a neutral, and a vertical part where
$g$ is Hermitian, $a$ anti-Hermitian, and $x_0$ satisfies (\ref{bu04}).
With the extra conditions

\begin{equation} \label{bu06}
\langle \phi_j, g \phi'_k \rangle =
\langle \phi_j, a \phi'_k \rangle = 0\,, \quad \, k, j  \geq n_w\,,
\end{equation}

both, $g$ and $a$, are unique. The conditions (\ref{bu06})
are equivalent to the choice of maximal null-spaces, i.e.
{\it minimal supports} for $g$  and $a$.
They allow to define $g$ and $a$ uniquely.

The transformation property (\ref{g01}) implies

\begin{equation} \label{g02}
w \mapsto w'=wu \, \Longrightarrow \, a \mapsto a' =
u^* a u + u^* \dot u\,,
\end{equation}

so that $x \mapsto a$ is a connection form (gauge potential) $\ba$ for
the gauge group $u \mapsto R_u$. However, support properties may not
change continuously. For parameter values at which the rank of $w$ is
changing, one has to understand $g$ or $a$ as equivalence class with
respect to the kernel of $g \mapsto gw$ or $a \mapsto wa$ respectively.
Then (\ref{g02}) remains meaningful even in those cases.

In our next step we look at $g$ and $a$. $g$, which describes the
horizontal part of a tangent vector $x$, can be expressed by
$\xi := \Pi_*x$ and $\varrho = ww^* := \Pi w$. We need
the pair $x$ and $\tilde \varrho := w^*w$ to gain $a$. We get

\begin{equation} \label{bu08}
\varrho \, g + g \, \varrho = \xi, \quad \tilde \varrho \, a +
a \, \tilde \varrho = w^*  x - x^* w\,.
\end{equation}

The first equation (\cite{Uh89b}, \cite{DG90})
is obtained from (\ref{tn01}). To see the second
one (\cite{Uh91a}), insert (\ref{bu05}) into its right hand side.

Apart from an obvious
restriction on $\xi$, (\ref{bu08}) can be solved to get $g$ or $a$,
and several ways to do so are well known. A review
is in \cite{BR97} .  The restriction in question reads
$\langle \phi, \xi \phi \rangle = 0$ whenever $\phi$ is in the
null space of $\varrho$ for the first equation, and
$\langle \phi', \xi \phi' \rangle = 0$ whenever $\phi'$ is in the
null space of $\tilde \varrho$. Below we assume they are satisfied.

With the solvability conditions in mind we rewrite (\ref{bu08}) as
equations between operators in $\cB(\cW)$.
In order not to overload notations we abbreviate

\begin{equation} \label{defLR}
\rL  \equiv L_\varrho, \, \, \rR \equiv R_\varrho, \, \,
\tilde \rL  \equiv L_{\tilde \varrho}, \, \, \tilde \rR \equiv
R_{\tilde \varrho} \,.
\end{equation}

These are families of operators indexed by
$\varrho$ or $\tilde \varrho$.

Let us start now from (\ref{bu08}). The equations can be solved  by

\begin{equation} \label{bu09}
g = ( \rL  + \rR )^{-1} \xi, \quad a = ( \tilde \rL  + \tilde \rR )^{-1}
( w^*  x - x^* w )\,.
\end{equation}

The operational defined inverse exists by the solvability condition
above. With two tangents $\xi_{j}$ at $\varrho$
and their horizontal lifts $x_j^{\rm hor}$ we get the Riemannian
metric, \cite{Uh92a},
\cite{Uh92b}, belonging to the Bures distance

\begin{equation} \label{bu10}
(\xi_2, \xi_1)^{\rm Bures} := \Re (x_1^{hor}, x_2^{hor})
= {1 \over 2} {\rm Tr} \, \varrho (g_1 g_2 + g_2 g_1)
\end{equation}

or, equivalently,

\begin{equation} \label{bu11}
(\xi_2, \xi_1)^{\rm Bures} = {1 \over 2} {\rm Tr} \, \xi_2 g_1
= {1 \over 2} {\rm Tr} \, \xi_2 ( \rL  + \rR )^{-1} \xi_1\,.
\end{equation}

There is a similar procedure with the second equation of (\ref{bu09})
resulting in the connection $\ba^{\rm geo}$ with  $\ba^{\rm geo}(x):= wa$.
The superscript
"geo", if used, is a reminder for the physical important
geometric phase. From (\ref{bu09}) we get

\begin{equation} \label{bu12}
\ba^{\rm geo} = {\tilde \rL  \over \tilde \rL  + \tilde \rR} ( w^{-1}
{\rm d}w) -
{\tilde \rR \over \tilde \rL  + \tilde \rR} ( w^{-1} {\rm d}w)^*\,,
\end{equation}

where $w^{-1} {\rm d} w$ is the left canonical 1-form
with values in the Lie algebra of GL$(\cH)$.
$\ba^{\rm geo}$ takes values in the Lie algebra of the
gauge  group U$(\cH)$ acting from the right  via $u\mapsto R_u$.

Formula (\ref{bu12})
represents $\ba^{\rm geo}$  as the difference of two Hermitian
conjugated parts of type (1,0) and (0,1) respectively:
$$
\ba^{\rm geo} = \ba_{1,0} - \ba_{0,1}, \quad \ba_{0,1} = \ba_{1,0}^*\,.
$$
Another interesting equation expresses $\ba^{\rm geo}$ as sum of the
canonical 1-form $\ba^{\rm can}$ of the bundle GL$(\cH)/$U$(\cH)$
and an horizontal Ad-1-form, \cite{DR92a},

\begin{equation} \label{bu13}
\ba^{\rm geo} = {w^{-1}{\rm d}w - (w^{-1}{\rm d}w)^* \over 2} +
{\tilde \rL  - \tilde \rR \over \tilde \rL  + \tilde \rR}
{w^{-1}{\rm d}w + (w^{-1}{\rm d}w)^* \over 2}\,.
\end{equation}

Since the second form is horizontal, it  can be rewritten in terms of
${\rm d}\varrho$ and we get

\begin{eqnarray} \label{bu14}
\ba^{\rm geo} &=& \ba^{\rm can} +
w^{-1} \bigl( {\rL  - \rR \over 2( \rL  + \rR)}
\,{\rm d}\varrho \bigr) ( w^{-1} )^*\\
&=&\label{bu14a}
w^{-1}{\rm d}w-
w^{-1}\bigl({ \rR \over \rL  + \rR}
\,{\rm d}\varrho \bigr) ( w^{-1} )^*   \,.
\end{eqnarray}

It becomes immediately clear that $\ba^{\rm geo}(x) = \ba^{\rm can}(x)$ iff
$\rL  \xi = \rR \xi$, where $\xi:=wx^*+xw^*$, i.~e. iff $\varrho$
commutes with $\dot \varrho$.

This observation motivates the decomposition
\begin{equation} \label{bu15}
\cT_{\varrho} =
\cT_{\varrho}^\para +
\cT_{\varrho}^{\bot}
\end{equation}
of the tangent space $\cT_{\varrho}$ into a direct sum,
where $\xi \in \cT_{\varrho}^{\para}$ iff
$\xi$ commutes with $\varrho = ww^*$ or, equivalently, iff
$\langle \phi_j, \xi \phi_k \rangle = 0$ for any two eigenvectors
$\phi_j$, $\phi_k$, of $\varrho$ with different eigenvalues.
On the other hand, $\xi \in \cT_{\varrho}^{\bot}$ iff it
can be written as a commutator  $i [b, \varrho]$ with a suitable
Hermitian $b$.
(\ref{bu15}) is a well known matrix decomposition: Assume $\varrho$
represented as block diagonal matrix, every block belongs to just
one eigenvalue. This induces a block representation of any matrix
$\xi$. One gets $\xi^{\para}$ by setting zero every off-diagonal
block of $\xi$. If the entries in the diagonal blocks are set
to zero, one obtains $\xi^{\bot}$. In our present field of interest
H\"ubner, \cite{Hu93a}, obtained a decomposition (\ref{bu15}) of
the Bures Riemannian metric. For larger classes of metrics this
has been done by Hasegawa and Petz (\cite{Ha93} \cite{PH95a}).

This brings us back to the metric (\ref{bu10}), (\ref{bu11}).
There is a solution
$g_1$ commuting with $\varrho$ iff $\xi_1$ does so: The
support $\varrho$ cannot be smaller than the support of $\xi$. Hence
$2g_1 = \varrho^{-1} \xi_1 = \xi_1 \varrho^{-1}$
is operational well defined.
Inserting in (\ref{bu11}) results in

\begin{equation} \label{bu17}
(\xi_2, \xi_1)^{\rm Bures} =
{1 \over 4} {\rm Tr} \, \xi_2 \xi_1 \varrho^{-1},
\quad \xi_1 \in \cT_\varrho^{\para}\,.
\end{equation}

Comparing this with the Riemannian metric

\begin{equation} \label{bu18}
(\xi_2, \xi_1)^{\rm can} := {1 \over 8} {\rm Tr} \,
( \xi_2 \xi_1 + \xi_1 \xi_2 ) \varrho^{-1} =
{\rm Tr} \, \xi_2 ( \rL ^{-1} + \rR^{-1} ) \xi_1
\end{equation}

the inequality $4/(\rL +\rR) \leq (1/\rL ) + (1/\rR)$ gives,
\cite{Pe95},

\begin{equation} \label{bu19}
(\xi, \xi)^{\rm Bures} \leq (\xi, \xi)^{\rm can}\,
\end{equation}

and equality holds if and only if $\xi \in \cT_{\varrho}^{\para}$,
or, what is the same, if $\xi$ commutes with $\varrho$.

Let $\phi_1, \dots $ a complete orthonormal eigenvector basis
of $\varrho = ww^*$ and $\xi$ with eigenvalues $\lambda_j$ and
$\dot \lambda_j$ respectively. Then we get from (\ref{bu17})
the following quadratic form
$$
 {1 \over 4} \, \sum {\rm d} \lambda_j^2 \, \lambda_j^{-1}
= \sum {\rm d} \mu_j^2, \quad \mu_j := \sqrt{\lambda_j}\,.
$$
This is an Euclidean metric. However, restricted to the state space,
where $\lambda_1, \dots$ becomes a probability vector, we get
Fisher's metric (``Fisher-Rao metric'') \cite{Fi25}.

{\it If the Bures metric is
restricted to a submanifold of mutual commuting states,
the Fisher metric is obtained.}

Moreover, {\it on any submanifold of commuting density operators},
whether normalized or not,
{\it the phase transport is holonomically trivial.}

Indeed, we can form the lift $\varrho \to w = \sqrt{\varrho}$.
The assumed commutativity
provides us with Hermitian and commutative $w$ and $x = \dot w$,
and with $\varrho =ww^* = w^*w = \tilde \varrho$.
Hence (\ref{bu09}) comes down to $\ba(x)=0$, and the lift is horizontal.
There is no room for a non-trivial phase.

We see, a non-trivial geometric phase is definitely an effect of
non-commutativity. We need for them curves with mutually not
commuting density operators.

\section{Auxiliary Tools}

In order to extend our previous considerations to a large the class
of connections, \cite{DR92a}, we need some auxiliary tools.

Looking at equations as (\ref{bu12}) or (\ref{bu14}) one can identify
functions of $\rL / \rR$ and $\tilde \rL / \tilde \rR$. These
operators are relatives of $\rL / \tilde \rR = \Delta_w$, the
Tomita-Takesaki modular operator of the representation $b \mapsto L_b$
with GNS-vector $w$.
The operators are defined if $w^{-1}$ exists, that is for completely
entangled $w$. But, as (\ref{bu12}) to (\ref{bu14}) show,
certain functions of these operators can be defined for every $w$.

Let $t \mapsto f(t)$ be a function defined for $0 < t < \infty$.
We assume the existence of

\begin{equation} \label{fu01}
f(0) :=  \lim_{t \to 0} f(t), \quad
f(\infty) :=  \lim_{t \to \infty} f(t)\,.
\end{equation}

The assumption is necessary if we like to extend the formalism
to density operators which are not invertible. Without it, we
have to restrict ourselves to completely entangled $w$,
i.e.~to faithful density operators.

To treat an example with the assumption (\ref{fu01}),
we define $f(\rL/ \tilde \rR) =: f(\Delta)$. The positive
operators $\rL$ and $\tilde \rR$ commute. Let $\lambda_j$ be the
eigenvalue of $ww^*$ and of $w^*w$ with the eigenvectors $\phi_j$ and
$\phi'_j$. The eigenvectors, suitably choosen, collect
in a complete orthonormal basis satisfying the Gram-Schmidt
decomposition (\ref{deco2}). $\lambda_j$ is zero if $j > n_w$ and positive
otherwise. Now

\begin{equation} \label{fu02}
\rL v_{jk} = \lambda_j v_{jk}, \quad \tilde \rR v_{jk} = \lambda_k
v_{jk},\quad v_{jk} := |\phi_j\rangle\langle\phi'_k|\,.
\end{equation}

The elements $v_{jk}$ constitute a complete orthonormal basis
of the Hilbert-Schmidt space $\cW$.
We like $f(\Delta)$ to be diagonalizable with eigenvectors
$v_{jk}$.  Remembering $\Delta = \rL / \tilde \rR$ we start with

\begin{equation} \label{fu03}
f(\Delta) \, v_{jk} = f(\lambda_j / \lambda_k) \, v_{jk}, \quad
{\rm if} \quad \lambda_k > 0\,.
\end{equation}

The remaining possibility is done ``by hand'' in requiring

\begin{equation} \label{fu04a}
f(\Delta) \, v_{jk} = f(\infty) \, v_{jk}, \, \,
{\rm if} \, \, \lambda_j > 0, \, \, \lambda_k = 0
\end{equation}

\begin{equation} \label{fu04b}
f(\Delta) \, v_{jk} = f(1) \, v_{jk},  \, \,
{\rm if} \, \, \lambda_j = \lambda_k = 0 \,.
\end{equation}

With this convention $v_{jj}$ is an eigenvector of $f(\Delta)$ with
eigenvalue $f(1)$ for all $j$.

The same game is to play with $f(\rL/\rR)$ and $f(\tilde \rL / \tilde
\rR)$. While the spectra of $f(\rL/\rR)$ and
$f(\tilde \rL / \tilde \rR)$ coincide with that of $f(\Delta)$,
their eigenvectors are, respectively,

\begin{equation} \label{fu05}
|\phi_j \rangle \langle \phi_k| = v_{ji} v_{ik}^*
\quad {\rm and} \quad
|\phi'_j \rangle \langle \phi'_k| = v_{ji}^* v_{ik}\,.
\end{equation}

\section{A Class of Connections }

Our aim is to describe a class of connections, essential that
of Dittmann and Rudolph, \cite{DR92a}.
These objects, as will be seen, are particularly
well adapted to the purification
of the $\cH$-system by that of $\cW = \cH \otimes \cH^*$.
We assume $w$ completely entangled,
so that $\varrho = \Pi w$ is faithful (invertible). Wether it is
possible to skip this assumption, either by calculating modulo neutral
tangents or by continuity arguments, depends on the asymptotic
behaviour of certain functions to be introduced below.

Let $[0,\infty] \ni s \mapsto r(s) \in \bbbc$ be a smooth function
and $r(1)=1/2$. Then

\begin{equation} \label{con01}
\left( r(\tilde \rL / \tilde \rR) y \right)^* = \bar
r(\tilde \rR / \tilde \rL) y^*\,.
\end{equation}

We get, therefore, a mimicked equation (\ref{bu12}) by

\begin{equation} \label{con02}
\ba :=
\bar r(\tilde \rL / \tilde \rR) ( w^{-1} {\rm d}w) -
r(\tilde \rR / \tilde \rL) ( w^{-1} {\rm d}w)^*\,.
\end{equation}

To transform like a connection it must have the form (\ref{bu13}),
though with an arbitrary horizontal Ad-1-form. Thus we need to have

\begin{equation} \label{con03}
\bar r(t) + r(1/t) = 1, \quad
F(t) := \bar r(t) - r(1/t) = - \bar F(1/t)
\end{equation}

to get a genuine connection with respect to the  gauge group U$(\cH)$
acting by $u \mapsto R_u$. Furthermore, as a consequence of
(\ref{con02}) and
$r(1)=1/2$, one observes rescaling invariance of the connection
form. Indeed, $\ba$ is invariant under
$w\mapsto\lambda(w)w$, where
$\lambda:\cW\rightarrow\bbbr$:

\begin{equation} \label{scale}
\ba_w(x)=\ba_{\lambda w}({\rm d}\lambda\,(x) \,w + \lambda x)
\end{equation}

so that there is no need to normalize $w$ in calculating $\ba$.
The second equation in (\ref{con03}) introduces the function
$F$ used in \cite{DR92a} to label their gauge potentials, and
we are allowed now to rewrite (\ref{con02}) in a manner
already known from (\ref{bu13}):

\begin{equation} \label{con04}
\ba = \ba^{\rm can} + F(\tilde \rL / \tilde \rR) \,
{(w^{-1}{\rm d}w) + (w^{-1}{\rm d}w)^* \over 2}\,.
\end{equation}

One returns to the Bures case by

\begin{equation} \label{conf}
\ba = \ba^{\rm geo} \Longleftrightarrow  r(t) = {t \over 1 + t}
\Longleftrightarrow F(t) = (t-1)/(t+1)\,.
\end{equation}

We may now proceed as in (\ref{bu14}) to get the deviation from
the connection $\ba^{\rm can}$. One obtains

\begin{equation} \label{con05}
\ba = \ba^{\rm can} +
w^{-1} \bigl( \, F(\rL/\rR) \,{\rm d}\varrho \, \bigr) ( w^{-1} )^*\,.
\end{equation}

Before deriving expressions  for the vertical and horizontal
part of a given tangent $x$, we draw an important conclusion:

{\it The value of a connection at the lift of a
$^{\scriptscriptstyle\|}$-tangent is independent of $F$
respectively $r$.}

Indeed, $F(1)=0$ and $\rL x = \rR x$ for these tangents, and
we get from (\ref{con05}) immediately

\begin{equation} \label{con06}
\Pi_*(x) \in \cT^{\para} \, \Longrightarrow \,
\ba(x) = \ba^{\rm can}(x), \quad \forall \, F
\end{equation}

allowing to extend a conclusion of section 4:

{\it On submanifolds with mutually commuting
density operators the holonomy of every loop is trivial for the
whole class of connections} considered here.

Indeed, the lift $\varrho \to \sqrt{\varrho}$ is horizontal
along every curve of commuting densities.

Now let us return to (\ref{con02}) and let us multiply this
equation by $w$ from the left. We obtain

$$
w\,\ba(x) =
\bar r( \Delta )(x) -
r(\Delta^{-1}) ( w x^* w^{* -1} )
$$

and, by the help of (\ref{con03}),

\begin{equation} \label{con07}
x^{\rm Ver} = w\,\ba(x) = x - r(\Delta^{-1}) ( x + w x^* w^{* -1} )\,.
\end{equation}

Reminding (\ref{deco3}) and (\ref{modcon1}), this can be seen
by the aid of the identities

$$
v^* (w^*)^{-1} = v^* (ww^*)^{-1/2} v = (w^*w)^{-1/2}
$$

$$
w x^* (w^*)^{-1} = \Delta^{1/2} J x = J \Delta^{-1/2} x\,.
$$

Another interesting equation, similar to (\ref{bu14a}), is

\begin{equation} \label{con08}
x^{\rm Ver} = x - \bigr( r( \rR/\rL ) \, \xi \bigr) (w^*)^{-1} \,.
\end{equation}

We assumed $w$ separating so that there are no non-vanishing
neutral tangents, and $x^{\rm Ver}=x^{\rm ver}$. Hence (\ref{con08})
or, equal well, (\ref{con07})
reflects the decomposition of a general tangent into a vertical and
an horizontal part, see (\ref{tnz}). We conclude

\begin{equation} \label{con09}
x^{\rm hor} = \bigr( r( \rR/\rL ) \, \xi \bigr) (w^*)^{-1}
= r( \Delta^{-1} ) \, [ x + \Delta^{1/2} J \, x ]\,.
\end{equation}

A connection form $\ba$ regulates the change of the phase $v$ along
an horizontal lift, $w_t=\sqrt{\varrho_t} \,v_t$, of a
curve $\varrho_t$. We express $\ba$ by

\begin{eqnarray}
\ba(\dot w)&=&
\ba(\sqrt{\varrho} \, \dot v+(\sqrt{\varrho})^\cdot\,v)=
\ba(\sqrt{\varrho} \,v \,v^* \dot v+(\sqrt{\varrho})^\cdot\,v)
=v^*\dot v+v^*\,\ba({\sqrt{\varrho}}\,^\cdot)\,v\nonumber\\
&=&v^*\dot v+v^*\,\ba
(\,\frac{1}{\sqrt{\rL}+\sqrt{\rR}}\,\dot\varrho\,)
\,v\nonumber\\
&=&
v^* \dot v + v^* \, {1 \over 2} {1 \over \sqrt{\rL\rR}}
\left( F(\rL/\rR)  + { \sqrt{\rR} - \sqrt{\rL} \over \sqrt{\rR} +
\sqrt{\rL}} \right)(\dot \varrho) \, v\,.
\end{eqnarray}

and see that horizontality of $w_t$ is equivalent with

\begin{equation} \label{con12}
0 = \dot v \, v^* + {1 \over 2} {1 \over \sqrt{\rL \rR}}
\left( F(\rL/\rR)  + { \sqrt{\rR} - \sqrt{\rL} \over \sqrt{\rR} +
\sqrt{\rL}} \right)\, (\dot\varrho)\,.
\end{equation}

One observes, that there is one and only one  connection in our setting
with a global horizontal section,
$\varrho\mapsto\sqrt{\varrho}$.
That connection is given by
$$
F(t) = - {1 - \sqrt{t} \over 1 + \sqrt{t}},\qquad
r(t) = {\sqrt{t} \over 1 + \sqrt{t}}.
$$

\section{Connection and Metric}

In this section we specify a class of Hermitian metrics
(\ref{tn03}) on  $\cW$,
which respects the purification
scheme. Our first task is to ask for Hermitian metrics
on the complex manifold $\cW$, the real
part of which is compatible with a given connection form of
the preceding section. We demand:
At every completely entangled $w \in \cW$, the vertical tangents
are real orthogonal to the horizontal ones.
In the case,
there exists an Hermitian metric doing this task,
the functions $F$ and $r$ charcterizing
the connection, have to be real.
In the next step we describe the Hermitian
an Riemannian metric one obtains by reduction from the
purification space to that of (unnormalized) density operators.

Starting with a connection (\ref{con02}), (\ref{con03}),
there is some freedom in the choice of the Hermitian metric.
It is an interesting question in its own, whether, by a
reasonable condition, the Hermitian metric becomes unique.
We explain in the last part of this section how this can be done.
If we start from a Riemannian metric on the density
operators, the  uniqueness problem is more involved.
Nevertheless, our additional condition
solves it also, at least for the monotone Riemannian metrics.

To start our little programm we construct Hermitian metrics
(\ref{tn03}) by modifying the Hilbert Schmidt scalar product
on $\cW$
by a function $k(\Delta)$ of the modular
operator. Like $\rR$ and $\rL$ the modular operator
$\Delta $ depends on $w$.
 Our
ansatz for the Hermitian product in ${\rm T}_w\cW$ reads

\begin{equation} \label{m01}
(x_2, x_1)_w := (x_2, k(\Delta_w)^{-1} x_1),
\end{equation}

where $k$ is a real positive smooth function defined either only on
$0 < t < \infty$ or on the closed interval $0 \leq t \leq \infty$.
We use the rules explained in the section ``auxiliary tools''.
There are two main merits with such a choice of the modified Hermitian
metric: The symmetry group of the metric contains the unitary group
$U(\cH) \times U(\cH^*)$. The second is the rescaling invariance of
$\Delta$ under $w \mapsto \lambda(w) w$, where
$\lambda(w)$ denotes (a sufficiently smooth) real function on $\cW$.
Rescaling invariance is a further reason not to insist in normalized
density operators.

In determining the connection form compatible with (\ref{m01}),
we follow the recipe of section 3. We need the real-orthogonal
complement of the vertical directions. They are
to gain by the metrical independence of verticality.
Namely, if a tangent $x$ is real-orthogonal to all vertical ones,
$k(\Delta)^{-1} x^{\rm hor}$ is horizontal w.~r.~to the
Hilbert-Schmidt-metric. Therefore,
as shown in section 4, we are allowed to write $x = gw$
with an Hermitian $g$. Conclusion:

{\it A tangent $x$ is horizontal with respect to (\ref{m01}),
iff it can be represented as}

\begin{equation} \label{mh1}
x = k(\Delta) (gw) = k(\rL/\rR)(g) w, \quad g = g^*\,.
\end{equation}

The real space of horizontal tangents is the fix point set of
an antilinear operator, $S_w^{k}$, acting on $\cW$.
Our notation is borrowed from that of the Tomita-Takesaki
operator $S_w = J \sqrt{\Delta}$, which will be returned if
$k \equiv 1$. Our definition is

\begin{equation} \label{mh2}
S_w^{k} = J k(\Delta^{-1}) k(\Delta)^{-1} \sqrt{\Delta} =
k(\Delta) k(\Delta^{-1})^{-1} S_w
\end{equation}

If this operator acts on $x = k(\Delta)(gw)$ the result is
$k(\Delta)(g^*w)$. Comparation with (\ref{mh1}) establishes:
$x$ is a fix point of $S_w^k$ if and only if $x$ is horizontal.

The square of the operator (\ref{mh2}) is $J^2$,
compare (\ref{modcon2}). $J^2$ is the identity of $\cW$ iff $w$
is invertible. Further, the adjoint of $S_w^k$ with respect to (\ref{m01})
is $\sqrt{\Delta}J$ and, as it should be, independent of $k$.
(Tomita-Takesaki theory calls it ``$F_w$''.) Finally we
polar decompose (\ref{mh2}) to get the appropriate modifications
of the modular operator, $\Delta = \Delta_w$, and of the modular
conjugation, $J = J_w$.

\begin{equation} \label{mh3}
S_w^k = J_w^k   |S_w^k, \quad \Delta_w^k := |S_w^k|^2,
\end{equation}

$$
\Delta_w^k = k(\Delta^{-1}) k(\Delta)^{-1} \Delta, \quad
J_w^k = J \sqrt{k(\Delta^{-1}) k(\Delta)^{-1}}
$$

We now ask for the connection comming with the metric.
The connection form belonging to (\ref{m01}) annihilates
all the horizontal vectors (\ref{mh1}).  This reasoning,
applied to (\ref{con02}) or (\ref{con04}), determines
the function $r$ or $F$. The calculation shows, in accordance
with (\ref{con03}),

\begin{equation} \label{m03a}
r(t) = {t \,k(1/t) \over k(t) + t \,k(1/t)}, \quad\mbox{ resp. }\quad
F(t) = { t \,k(1/t)-k(t) \over  t \,k(1/t)+k(t)}\,.
\end{equation}

Obviously, {\it the functions $r$ and $F$ are real valued} if the
connection is gained from an Hermitian metric (\ref{m01}).
A cross check of (\ref{m03a}) is in setting $k \equiv 1$. We
get $r(t) = t/(1+t)$ and $F(t)=(t-1)/(t+1)$ as it should be
for the Bures case.

On the
other hand, given $r$ or $F$, there is some freedom for $k$
since the induced connection depends on $k(t)/k(1/t)$ only.

\begin{eqnarray*}
{k(t) \over k(1/t)} = 1 & \Longleftrightarrow& r(t) = {t \over 1 + t},
\quad F(t) = {(t-1) \over (t+1)}, \quad
\ba = \ba^{\rm geo}\\
{k(t) \over k(1/t)} = t & \Longleftrightarrow&  r(t) = {1\over 2},
\qquad\, F(t)=0,\quad
\ba = \ba^{\rm can}
\end{eqnarray*}

In particular, there is no modification of the Tomita-Takesaki
operators by (\ref{mh2}), (\ref{mh3}) if the connection
is $\ba^{\rm geo}$.
More generally, from (\ref{m03a}) we get

\begin{equation}\label{kF}
{k(t) \over k(1/t)} = t {r(1/t) \over
r(t)}=t{1-F(t)\over1+F(t)}
\end{equation}

and find, remarkably enough, the modified Tomita-Takesaki operators
(\ref{mh2}), (\ref{mh3}) depending on $F$ only. Further,
by (\ref{kF}), the positivity of $k$ enforces the inequality

\begin{equation}\label{ungl}
-1 < F(t) < 1
\end{equation}

for $F$ to be obtained from a $k$. In order to invert (\ref{kF}),
the inequality is also sufficient. (According to (\ref{con03}) one
needs only to check $F < 1$ for real $F$.) Then, given $F$,
the general solution of the problem is

\begin{equation}\label{qk}
k(t) :=\sqrt{t}\,(1-F(t))  \, q(t)\,,
\end{equation}

$q$ being an arbitrary positive function fulfilling $q(t)=q(1/t)$.

We started from an Hermitian metric on $\cW$, derived conditions
for horizontality, and determined the connection. Now we go back
to $\cH$ and to its density operators: We ask for the Hermitian
and Riemannian
metric induced on the space of density operators. That is,
with two tangents $\xi$ and $\eta$ at $\Pi w = \varrho$, we are
concerned with

\begin{equation} \label{m04}
(\eta, \xi)_{\varrho} := (y^{\rm hor}, x^{\rm hor})_w
\end{equation}
and
\begin{equation} \label{m04b}
\Re(\eta, \xi)_{\varrho} =
\frac{(\eta, \xi)_{\varrho} + (\xi, \eta)_{\varrho}}{2}\,.
\end{equation}
$x^{\rm hor}$ and $y^{\rm hor}$ are horizontal lifts of
$\xi$ and $\eta$. In the present paper the  $\bbbc$-valued $\bbbr$-linear
form (\ref{m04}) is defined on the real tangents. Nevertheless, for obvious
reasons, we call it
``Hermitian''. Relying on (\ref{con09}) we conclude

\begin{equation} \label{m05a}
(y^{\rm hor}, x^{\rm hor})_w =
{\rm Tr} \;  r(\rL/\rR)( \eta)\, \frac{r(\rR/\rL)}{\rR\,k(\rL/\rR)}(
\xi)=
{\rm Tr} \,  \eta \, {r(\rR/\rL)^2 \over \rR \, k(\rL/\rR)} \, \xi
\end{equation}

so that

\begin{equation} \label{m05}
(\eta, \xi)_{\varrho} =
{\rm Tr} \,  \eta \, {\rR \,k(\rL/\rR) \over [\rR \,k(\rL/\rR) + \rL
k(\rR/\rL)]^2} \, \xi\,,
\end{equation}

where $r$ has been substituted by $k$ in (\ref{m05a})
by the aid of (\ref{m03a}). The real part of (\ref{m05}) is
a Riemannian metric. By   (\ref{m05}) and standard rules
we get

\begin{equation} \label{rm}
\Re (\eta, \xi)_{\varrho} =
{1
\over 2} \,{\rm Tr} \, \eta \, { 1 \over \rR \,k(\rL/\rR) + \rL\,
k(\rR/\rL)} \, \xi\,.
\end{equation}
Petz, \cite{Pe95}, \cite{Pe96}, \cite{PS96},  was able to classify
all monotone Hermitian metrics on the state space, i.~e.~those for which
$( \cdot , \cdot )_{\varrho}$
does not increase under the action of completely positive and unital
mappings. On the heart of his result is the characterization
of a monotone metric by an operator monotone function, $f$,
defined on $0 < t < \infty$, such that

\begin{equation} \label{m07}
(\eta, \xi)_{\varrho} =
\frac{1}{4}\,{\rm Tr} \, \eta \; \frac{\rR^{-1}}{ f(\rL/\rR)}
\xi\,.
\end{equation}
(The factor $1/4$ is a normalization convention.)
Note, that this
Hermitian metric  becomes symmetric, and hence a
Riemannian one,
if and only if the function $f$ satisfies $f(t) = t f(1/t)$.
A function with this algebraic property we call selftransposed following
the terminology for operator means introduced in \cite{KA80}.
Presently,
however, the monotonicity of the metric (\ref{m07})
or of its real part is {\it not} assumed. We need a
more general frame. Having this in mind, we compare (\ref{m07})
with (\ref{m05}) and obtain

\begin{equation} \label{m08}
{f(t)} =
{ (\,k(t) + t k(1/t)\,)^2  \over 4\,k(t) }\,.
\end{equation}

This equation has a unique solution for $k$ depending on $f$, therefore,
every Hermitian metric (\ref{m07}) can be reached by exactly one Hermitian
metric  (\ref{m01}) on the purification space.
Indeed, the harmonic mean of $f(t)$ and its transpose, ${tf(1/t)}$,
yields

$$
{1 \over f(t)} + {1 \over tf(1/t)} = {4 \over k(t) + t k(1/t)}
$$

so that one can insert this into the right hand side of (\ref{m08}) to
express $k$ by $f$:

\begin{equation} \label{m09}
k(t) = f(t) \, {4t^2 f(1/t)^2 \over [f(t) + t \, f(1/t)]^2}\,.
\end{equation}

Moreover, using (\ref{m03a}) we get

\begin{equation} \label{m10}
r(t) = { f(t) \over f(t) + t \, f(1/t)}
\quad \mbox{and} \quad
F(t)={f(t)-t \,f(1/t)  \over f(t)+t \,f(1/t) }\,.
\end{equation}

These equations describe the relation between the connection on
$\cW$ and the Hermitian metric living on the density operators.
It is Riemannian iff $f$ is selftransposed.
(\ref{m09}) yields $f = k$ in this case, and (\ref{m10}) degenerates
to $r \equiv 1/2$. Hence, \newline{\it if the induced Hermitian form is
Riemannian, the induced connection is
necessarily the canonical one.} \newline
This way we do not get an interesting mapping from the class of
Riemannian metrics to the class of connections.
Especially, the function $f(t)=(1+t)/2$ belonging
to the Bures metric can not
be gained from $\ba^{\rm geo}$ as one might expect.

Moreover,
if we like to gain the connection form $\ba^{\rm geo}$, $r(t)=t/(t+1)$,
belonging to the geometric phase, we need, according to (\ref{m10}),
 $t^2 f(1/t) = f(t)$ or, equivalently, $k(t)=k(1/t)$.
If $f$ is operator monotone, so is $tf(1/t)$. Therefore,
$t^2f(1/t)$ is convex (lemma 5.2 of \cite{KA80}). Thus, $f$
is convex and, as an operator monotone function, concave.
Being convex and concave, $f$ it has to be affine.
An affine function on the positive real axis, fulfilling
$t^2 f(1/t) = f(t)$, is a multiple of $t$.

{\it If $\ba = \ba^{\rm geo}$ and $f$ is operator monotone with
$f(1) = 1$, then} $f(t) = t$.

However, considering the real part we obtain for $k(t)=1$
(resp.~$k(t)=2t/(t+1)$)  $\ba=\ba^{\rm geo}$ (resp.~$\ba=\ba^{\rm can}$)
 and
\begin{equation}\label{n1}\Re (\eta, \xi)_{\varrho}=
\frac{1}{4}\,{\rm Tr} \, \eta \, \frac{R^{-1}}{f_s(L/R)}\,\xi
\end{equation}
with
$f_s(t)=(1+t)/2$ (resp.~$f_s(t)=2t/(t+1)$).
These  $f_s$ are distinguished
(selftransposed) operator monotone
functions. Moreover, in these cases the real part of
the Hermitian metrics
(\ref{m01}) restricted to the horizontal vectors coincides with the real
part of the Hilbert-Schmidt metric. This is the motivation to deal in
the following with the real part of the Hermitian metric induced on the
state space.

First of all, this Riemannian metric  is of the form  (\ref{n1}) with a
certain
selftransposed function $f_s$ depending on $k$. From (\ref{rm}) we get
\begin{equation}\label{n2}
f_s(t)=\frac{k(t)+t\,k(1/t)}{2}\,.
\end{equation}
$f_s(t)$ is the harmonic mean of $f(t)$ and $t f(1/t)$, $f$ given by
(\ref{m08}).

Clearly, in starting with a selftransposed $f_s$
there is some arbitrariness
in choosing $k$ respecting (\ref{n2}).
Moreover, given a selftransposed $f_s$, the only  restriction for
$F$ is  $-F(1/t)=F(t)<1$. Indeed, the equations
(\ref{m03a}) and (\ref{n2}) then have the  unique solution
\begin{equation}\label{hfF}
k(t) = f_s(t) \, ( \, 1 - F(t) \, ) \,.
\end{equation}

In order to remove the arbitrariness in going from $f_s$ to $F$
and vice versa or from $f_s$ to $k$, we impose an additional requirement on
the
class (\ref{m01}) of Hermitian metrics $(x, y)_w$. The aim is
to ensure that, given $f_s$, there is only one $k$ and one $F$
fulfilling (\ref{m03a})  and (\ref{n2}). We
shall prove that we meet our goal for operator monotone $f_s$
by the following natural
demand:

{\it Condition HS : For $x$ and $y$ belonging to the
horizontal spaces
defined by the Hermitian metric (\ref{m01}), the real part,
$\Re (x, y)_w$, of $(x, y)_w$
coincides with the real part, $\Re (x, y)$,  of the Hilbert-Schmidt
product of $x$ and $y$. }

At first, by the aid of (\ref{mh1}), the condition HS becomes

\begin{equation}\label{beding1}
\Re\left(k(\Delta)(gw),g'w\right)=
\Re\left(k(\Delta)(gw),k(\Delta)(g'w)\right)
\end{equation}

with arbitrary Hermitian $g$ and $g'$. It yields the constraint

\begin{equation}\label{beding2}
k(t)+t\,k(1/t)=k(t)^2+t\,k(1/t)^2\,.
\end{equation}
Next, we have the following crucial observation, which one verifies
straightforwardly:
\newline {\it There is an one-to-one
correspondence beetween  positive functions $k$ fulfilling the constraint
(\ref{beding2})
and functions $F$ with $-F(1/t)=F(t)<1$ .
The correspondence
is given by (\ref{m03a}) and}
\begin{equation}\label{m03b}
k(t)=\frac{2t(1-F(t)}{(1+F(t))^2+t(1-F(t))^2}\,.
\end{equation}
By (\ref{n2}) or, equally well, by (\ref{hfF})
we get  the relation between $F$ and $f_s$
\begin{equation}\label{fF}
f_s(t)=\frac{2t}{(1+F(t))^2+t(1-F(t))^2}\,.
\end{equation}
Hence, under condition HS, a function
$f_s$ can be gained from a $k$ iff $f_s$ has a representation
(\ref{fF}) with a suitable $F$, $F(t)<1$.

To explain, which functions $f_s$ can be reached, we rewrite relation
(\ref{fF}) into the equivalent form

\begin{equation}\label{fFb}
\frac{1+t}{2}-f_s(t)=
\frac{f_s(1/t)\,(1+t)^2}{4}\,
\left(\,\frac{t-1}{t+1}-F(t)\,\right)^2\,.
\end{equation}

Therefore,  necessary conditions for $f_s$ are $f_s(1)=1$,
$f_s\leq(1+t)/2$ and, moreover, $t\mapsto (1+t)/2 -f_s(t)$
must be  the
square of a smooth function.

Now suppose, we have such a pair $f_s,F$.
We define  an auxiliary smooth function

$$\delta(t):=
\frac{\sqrt{f_s(1/t)}\,(1+t)}{2}\,
\left(\,\frac{t-1}{t+1}-F(t)\,\right)\,.
$$

It fulfils

\begin{equation}\label{delta}
\delta(t)^2 = \frac{1+t}{2}-f_s(t)
\,,\qquad \mbox{and }\quad
\sqrt{t} \, \delta(1/t) + \delta(t) = 0 \,.
\end{equation}

The second equation is a consequence of $F(1/t)=-F(t)$ and
$f_s(t)=t \,f_s(1/t)$.
$F$ can be expressed in terms of $\delta$ and $f_s$ by

\begin{equation}\label{sol}
F(t) = \frac{t-1}{t+1}-\frac{2}{(1+t)\,\sqrt{f_s(1/t)}} \, \delta(t) \,.
\end{equation}

Conversely,  for a given selftransposed $f_s$, $f_s(1)=1$,
the possibilities in choosing $\delta$ with the properties
(\ref{delta}) enumerate via (\ref{sol}) the solutions  $F$
of (\ref{fF}) and $-F(1/t)=F(t)$.
But such an $F$ may not fulfil $F(t)<1$ if we did not choose
appropriately the signs for $\delta$ in (\ref{delta}). The wanted
choice may be neither unique nor possible. But if so,
the function $k$ defined by

\begin{equation}\label{n3}
k(t):=\frac{2}{t+1}\,\left(f_s(t)+\sqrt{t\,f_s(t)}\,\delta(t)\right)
\end{equation}
satisfies (\ref{n2}) and (\ref{m03a}).

The question, which functions $f_s$, $f(1)=1$,
bounded by $0<f(t)\leq (1+t)/2$,
can arise from $F$ or, equivalently, from an Hermitian metric (\ref{m01}),
depends also on  regularity requirements on $F$ and $k$. We do not
discuss this  in detail.
Instead we have the following  uniqueness result:
\newline
{\it For every  selftransposed operator monotone function
$f:(0,\infty)\rightarrow\bbbr$
 with $f(1)=1$ there exists exactly one positive function $k$
fulfilling (\ref{n2}) and (\ref{beding2}).}

We prove this assertion in the Appendix. We will  also show, that
such a selftransposed $f_s$ is of the form
\begin{equation}\label{fs}
f_s(t)=\frac{1+t}{2}-(t-1)^2 \,\tau(t)^2 \,,
\end{equation}
where $\tau$ is a strictly positive  function or equal zero
and the corresponding functions $k$ and $F$ then are given by

\begin{eqnarray}\label{kf}
k(t)&=&
\frac{2f_s(t)}{1+t}\,\left(1+\frac{(t-1)\,\tau(t)}{\sqrt{f(1/t)}}\right)\\
\label{Ff}F(t)&=&
\frac{t-1}{t+1}\;\left(\,1-\frac{2\,\tau(t)}{\sqrt{f(1/t)}}\,\right)
\end{eqnarray}
Thus we get:

{\it For every  monotone Riemannian metric
(\ref{n1}), $f_s(1)=1$, on the manifold
of completely entangled states
 there exists exactly one Hermitian
metric (\ref{m01})  satisfying the condition HS
such that the real part of the induced Hermitian  metric is just the given
monotone metric. For $f_s$ given by (\ref{fs}) the Hermitian
metric and the corresponding connection form are obtained from
(\ref{kf}) and (\ref{Ff}.}

The obtained  connection we
call the connection associated to the monotone Riemannian metric.
For the Bures metric we return to the Hilbert-Schmidt metric and the
connection above called $\ba^{\rm geo}$.

Since  we used only certain properties of operator
monotone functions this assertion  would be true for a larger
class of metrics, but we will not deal with this problem.

Although the condition HS seems to be  natural, perhaps a short comment
would be worthwhile. The induced Riemannian metrics are obtained,
essentially, by taking the real part of the Hermitian metric of
horizontally lifted vectors. But, because of HS, this is the same as the
real part of the Hilbert-Schmidt metric. Forgetting for a moment about
the underlying Hermitian metric, which forced horizontality,
we can take the following point of view: The monotone metrics are
obtained from the originally given Hilbert-Schmidt metric similary to
the Bures metric (section 4). The
deviation from the Bures metric is caused by some constraints
on the purifying lifts.

\section{Examples}

At first we look at curves of density operators
satisfying a von Neumann equation

\begin{equation} \label{hamilton1}
i \dot \varrho = [h, \varrho], \quad h = h^*, \quad \dot h = 0
\end{equation}

and their lifts. We may think of $h \in \cB(\cH)$ as of a given Hamiltonian
and of the curve parameter, $t$, as time. This interpretation is not
obligatory: $h$ may be the generator of any one-parameter group.
(The parameter $t$ should not be confused with the use
of the same letter as a dummy variable in several functions
like $f$, $k$, $r$, $F$.) To fix a solution of (\ref{hamilton1}),
we start at an initial time, $t_{in}$, with an initial density
operator $\varrho_{in}$. The solution may be written

\begin{equation} \label{hamilton2}
\varrho_t = u_t^* \varrho_{in} u_t, \quad u_t := \exp i (t - t_{in}) h
\,.
\end{equation}

Now a general lift $w_t$ is polar decomposed,
$w_t = \sqrt{\varrho_t}v_t$, according to (\ref{deco3}).

Our aim is to prove the following: {\it Given a connection form
and an initial $\varrho_{in}$ at $t_{in}$. There is a t-independent
Hermitian $\tilde h$ such that}

\begin{equation} \label{hamilton4}
u_t v_t = \exp i (t - t_{in}) \tilde h
\end{equation}

{\it implies horizontality of $w_t$.}

At first we see from (\ref{hamilton2}) and (\ref{hamilton4})
the validity of a Schr\"odinger equation in $\cW$,

\begin{equation} \label{hamilton5}
i \dot w = H w, \quad H w := h w - w \tilde h
\,.
\end{equation}

By the help of our menagerie of equations it is not particular
difficult to prove the statement above and to obtain an expression
for $\tilde h$. At first let us multiply (\ref{hamilton5}) by
$w^*$ from the right. By (\ref{con09}) the condition for
horizontality is in equating $i \dot w w^*$ with
$r(R/L)i \dot \varrho$. Now (\ref{hamilton1}) yields

\begin{equation} \label{hamilton6}
r(\rR / \rL) (h \varrho - \varrho h) = h \varrho - w \tilde h w^*
\,.\end{equation}

This equation is sufficient to guarantee horizontality.
Now $w\tilde hw^*$ can be computed by (\ref{hamilton4}) to
$u_t^* \sqrt{\varrho_{in}} \tilde h \sqrt{\varrho_{in}} u_t$.
Therefore, our horizontality condition is the Ad-transform
with $u_t^*$ of the equation

$$
r(\rR_{in} / \rL_{in}) (h \varrho_{in} - \varrho_{in} h) =
h \varrho_{in} - \sqrt{\varrho_{in}} \tilde h \sqrt{\varrho_{in}}
\,,
$$

where $\rR$ and $\rL$ at $t=t_{in}$ is indexed by {\it in}.
In other words, if we choose $\tilde h$ t-independent and $v$ according
to (\ref{hamilton4}), we can satisfy the horizontality condition.

To get a unique $\tilde h$, we require the support of $\tilde h$
to be smaller than that of $\varrho_{in}$. Finally, by the help of
(\ref{con03}), we get the expression

\begin{equation} \label{hamilton7}
\tilde h = \Bigl( \sqrt{\rR / \rL} \, r(\rL / \rR) +
\sqrt{\rL / \rR} \, r(\rR / \rL) \Bigr) \, h, \quad t = t_{in}
\,.
\end{equation}

Let us consider a solution (\ref{hamilton2}) of (\ref{hamilton1})
from $t_{in}$ to $t_{out}$. {\it Then $w_{out}w_{in}^*$ is a gauge
invariant.} Its trace in $\cH$,
\begin{eqnarray}
( w_{in}, w_{out} ) &=& ( w_{in}, [ \exp i(t_{out} - t_{in})H]  w_{in}
)\nonumber\\
&=& \label{hamilton8}
{\rm Tr} \, \sqrt{\varrho_{in}}\sqrt{ \varrho_{out}}\,
\exp(i(t_{in} - t_{out}) h)\,
\exp( i(t_{out} - t_{in}) \tilde h)\,,
\end{eqnarray}
may be called a {\it relative geometric phase}. For pure states
that object has been introduced in \cite{SB88}. These authors called it
``non-cyclic geometric phase''. One may think of shortcutting
the {\it in}- and the {\it out}-state to a closed curve by a
Fubini Study geodesic arc. Whether one has a similar interpretation
in our much more general case remains an open question.

For a cyclic solution of (\ref{hamilton1}), i.e.
$\varrho_{in} = \varrho_{out}$, $t_{cycle} = t_{out}-t_{in}$,
the expression $w_{out}w_{in}^*$ is a (pointed) holonomy invariant,
i.e. it depends on the choice of $\varrho_{in}$. To change the
{\it in}-state of our cyclic curve
one has to perform a $u_t$-transformation.
Consequently, {\it all eigenvalues of $w_{out}w_{in}^*$  are
(absolute) holonomy invariants.}
of our cyclic curve. They are encoded in the traces

\begin{equation} \label{hamilton9}
{\rm Tr} \, (w_{out}w_{in}^*)^m = {\rm Tr} \,
[\varrho_{in}\,\exp (-i t_{cycle} h)\, \exp (i t_{cycle} \tilde
h)]^m\,,
\end{equation}
where  $\exp (-i t_{cycle} h)$ commutes with $\varrho_{in}$.

 \hspace{0.5cm}There are a few examples where on can become more explicit.
One of them is in {\it adding noise to a curve of
pure states} $p_t$. In this important example one can study the
influence of ``noise''
on the geometric phase, and the behavior
of gauge and holonomy invariants in coming from the interior
to the extreme boundary of the
cone of unnormalized density operators.
For this purpose we  fix two positive real numbers,
$\alpha$ and $\beta$, and consider the curve of density operators
$\varrho$
\begin{equation} \label{h01}
\varrho = \alpha \, p + \beta \, \bbbone, \quad
p = |\psi \rangle\langle \psi|, \quad \langle \psi, \psi \rangle
= 1\,.
\end{equation}
$\alpha + \beta$ is a simple and $\beta$, if $n$ denotes the
dimension of $\cal H$, a $(n-1)$-fold eigenvalue of $\varrho$.
$\psi$, $p$ and $\varrho$ depend  on a parameter
$t$, but we will not suppose a v. Neumann
equation.

{\bf Remark:} The line element of this curve w.~r.~to the metric induced from
(\ref{m01}) is
$${\rm d}s^2=\frac{2\alpha(1-\tau)}{\tau k(1/\tau)+k(\tau)}\,
{\rm d}s^2_{Bures}\,,\qquad\tau:=\frac{\beta}{\alpha+\beta}\,,$$
where ${\rm d}s^2_{Bures}$ denotes the Bures line element of the curve of
pure states
$p_t\,$.$\qquad\Box$

All $t$-derivations will be indicated by a dot,
in particular
\begin{equation} \label{h02}
\dot \varrho = \alpha \dot p, \quad \dot p = \dot p p
+ p \dot p, \quad p \dot p p = 0\,.
\end{equation}
$\dot \varrho$ belongs to $\cT^{\perp}$.
As an application one calculates
$$
R_{\varrho} \dot p = \dot p (\alpha p + \beta \bbbone)
= (\alpha + \beta) \dot p p + \beta p \dot p\,.
$$
In this manner one gets
\begin{eqnarray} \label{h03}
R_{\varrho} (p \dot p) = \beta p \dot p, \qquad\quad\;\;&&
R_{\varrho} (\dot p p) = (\alpha + \beta) \dot p p\,,\\
 \label{h04}
L_{\varrho} (p \dot p) = (\alpha + \beta) p \dot p, \quad&&
L_{\varrho} (\dot p p) = \beta \dot p p
\end{eqnarray}
and, finally, skipping the index of $L_{\varrho}$ and $R_{\varrho}$,
\begin{equation} \label{h05}
(\rL/\rR) (p \dot p) = ({\alpha + \beta \over \beta}) p \dot p,
\quad
(\rL/\rR) (\dot p p) = ({\beta  \over \alpha + \beta}) \dot p p\,.
\end{equation}
For instance, $\dot p p$ and $\dot p p$ are eigenvectors of
$\rL\rR$ with the eigenvalue $(\alpha + \beta) \beta$.
At this stage we do not suppose a von Neumann equation
(\ref{hamilton1}) but rely on (\ref{con12}). Reminding (\ref{h05})
and $F(t) = -F(1/t)$, we get
$$
F(\rL/\rR) \dot p =
F({\beta  \over \alpha + \beta}) ( \dot p p - p \dot p )\,.
$$
Hence, in solving (\ref{con12}) with (\ref{h01})
we are faced with an equation
\begin{equation} \label{tp04}
\dot v \, v^*  =
{1 \over 2} {\alpha \over \sqrt{(\alpha + \beta) \beta}} \Bigl[
F({\beta  \over \alpha + \beta}) + {\sqrt{\alpha + \beta} - \sqrt{\beta}
\over \sqrt{\alpha + \beta} + \sqrt{\beta}} \Bigr]
( \,p \, \dot p - \dot p \, p\, )\,,
\end{equation}
which may be rewritten as
\begin{equation} \label{tp05}
\dot v^*   = v^* (1 - \mu) (p \, \dot p - \dot p \, p),
\quad \mu =
{1 \over 2} {\alpha \over \sqrt{(\alpha + \beta) \beta}} \bigl[
F({\beta  \over \alpha + \beta}) + {\alpha + 2 \beta
\over \alpha} \bigr]\,.
\end{equation}
Can we go by $\beta \to 0$ to the pure states? A necessary
condition is
\begin{equation} \label{lim1}
F(0) = -1
\end{equation}
or, equivalently, $r(0) = 0$. To be sufficient we additionally need
the existence of
\begin{equation} \label{lim2}
\kappa := \lim_{\beta \to 0} \mu
= \lim_{\lambda \to 0} {1 + F(\lambda) \over 2 \sqrt{\lambda}}
=  \lim_{\lambda \to 0} \lambda^{-1/2} r(\lambda)\,.
\end{equation}

Then the limit $\beta \to 0$ can be performed in (\ref{tp04}):

\begin{equation} \label{lim3}
(v \, \dot v^*)^{\rm pure}  = (1 - \kappa)
(p \, \dot p - \dot p \, p)\,.
\end{equation}

With $a^{geo}$, or, more generally, with $s > 1/2$ in $r = \lambda^s/(1
+ \lambda^s)$, we get $\kappa = 0$. {\it With $\kappa = 0$ we obtain the
Berry phase for pure states.}

Indeed, imposing $\langle \psi, \dot \psi \rangle = 0$ a la
Berry \cite{Be84} and Fock \cite{Fo28}, we find
$\dot v^* \psi + v^*\dot \psi = 0$
from (\ref{tp05}). Hence, with $\kappa = 0$, the vector $v^* \psi$
is t-independent. This yields
$w = |\psi\rangle \langle \varphi|$, $\dot \varphi = 0$. It then
follows

$$
{\rm Tr} \, (w_{out}w_{in}^*)^m = \langle \psi_{in}, \psi_{out}\rangle^m
\,.
$$

This is the $m$-th power of the Berry phase,
because we had supposed the validity of Berry's transport condition.
Remark that this goes
not through if $\kappa \neq 0$ or if, as for $a^{can}$,
(\ref{lim2}) does not exist.

Something more can be said if (\ref{h01}) satisfies a von Neumann
equation (\ref{hamilton1}). Computing $\tilde h$ with this
assumptions by the help of (\ref{hamilton7}) ends up with

\begin{equation} \label{tp06}
\tilde h = h + \mu \, [ ( \bbbone - p_{in} ) h \,p_{in} +
p_{in} \,h ( \bbbone - p_{in} ) ]\,.
\end{equation}

Looking at $\tilde h$ as a block matrix with respect of $p_{in}$
and $\bbbone - p_{in}$, the deviation from $h$ is in
multiplying the off-diagonal blocks by $\mu$. If (\ref{lim2})
exists and $\kappa = 0$ then the off-diagonal blocks become zero
at the pure state limit.
\section{Appendix}
Every selftransposed operator monotone
function $f_s$ has  a unique integral representation

\begin{eqnarray}
f_s(t) &=&m
(\{0\})\,\frac{1+t}{2}+\int\limits_{(0,1]}\frac{1+x}{2}\left(\frac{t}{t+x}+
\frac{t}{t\, x+1}\right)\,{\rm d}m(x) \nonumber\\
&=& \frac{1+t}{2} +
\int\limits_{(0,1]}\left\{-\frac{1+t}{2}+\frac{1+x}{2}\left(\frac{t}{t+x}+
\frac{t}{t\, x+1}\right)\right\}\,{\rm d}m(x) \nonumber\\
&=& \frac{1+t}{2} - (1-t)^2 \,
\int\limits_{(0,1]}\frac{x(t+1)}{2(t+x)(t\,x+1)}\,{\rm d}m(x) \,,
\label{integralrep}
\end{eqnarray}

where $m$ is a normalized positive Radon measure  on $[0,1]$
(see \cite{KA80}).
If the measure is not concentrated at 0,  the last integral is
strictly positive for all $t\in\bbbr_+$. Its positive root,
for the time being denoted by $\tau$, is a real analytic
function. Hence,

\begin{equation}\label{fmono}
f_s(t) = \frac{1+t}{2}-(t-1)^2 \, \tau(t)^2 \,
\end{equation}

and $(1+t)/2-f_s(t)$ has exactly two real analytic roots,

$$
\delta_+(t)=(t-1)\,\tau(t) \qquad
\delta_-(t)=-(t-1)\,\tau(t) \,,
$$

or is vanishing.
The selftransposeness of $f_s$ implies $\tau(1/t)=\sqrt{t}\,\tau(t)$
and both roots fulfill the condition (\ref{delta}).
 From(\ref{sol}) we infer:
If selecting the root $\delta_+$, the condition $F(t)<1$, $t>0$,
is equivalent to $ f_s(t)>1/2$ for all $t>1$.
Because $f_s$ is monotone increasing and $f_s(1)=1$ the latter
inequality is true.
On other hand, $F$ can not fulfil $F(t)<1$ for all $t>1$
if the root
$\delta_-$ is chosen, except , $\delta_-= 0$.
Otherwise we could conclude
$f_s(t)>t/2$ for all $t>1$. But the selftransposeness effects
$f_s'(1)=1/2$ and $f_s$ must be concave.
Therefore, $\delta(t):=(t-1)\,\tau(t)$ is the only  root
leading to an appropriate $F$ and formulae (\ref{n3}), (\ref{sol})
yield (\ref{kf}), (\ref{Ff}).

{\bf \noindent Acknowledgement}
\nopagebreak

The authors like to thank P.M.~Alberti, C.~Crell, and M.B.~Ruskai
for valuable remarks. Our particular thank goes to D\`enes Petz
and Gerd Rudolph for enlightening and helpful discussions. J.~D.~ is grateful to
H.~B.~Rademacher for a valuable advice.
The authors gratefully acknowledge a stay at the
``Stefan Banach International Mathematical Center'', Warszawa, and
one of us (A.U.) likes to thank the ``Erwin Schr\"odinger International
Institute for Mathematical Physics'', Vienna, and the
``1997 Elsag-Bailey -- I.S.I. Foundation
research meeting on quantum computation'', Torino, where part of the
research has been done.

{\bf Postal addresses:}

\vspace{1.5cm}
\noindent
\begin{tabular}{lcl}
Jochen Dittmann&\hspace{9cm}&Armin Uhlmann\\
Mathematisches Institut&&Institut f.~Theoretische Physik\\
Universit\"at Leipzig&&Universit\"at Leipzig\\
Augustusplatz 10/11&&Augustusplatz 10/11\\
04109 Leipzig&&04109 Leipzig\\
Germany&&Germany
\end{tabular}

\vspace{2cm}
e-mail correspondence: {\tt dittmann@mathematik.uni-leipzig.de}

\begin{thebibliography}{**}
\bibitem{books1}
Geometric Phases in Physics. (ed. A.~Shapere, F.~Wilczek),
World Scientific Publishing Co., Singapure, 1990. \hfill \break
Anomalies, Phases, Defects (ed. M.~Bregola, G.~Marmo,
G.~Morandi), Bibliopolis, Naples, 1990. \hfill \break
Topological Phases in Quantum Theory. (ed.~B.~Markowski,
S.~I.~Vinitski), World Scientific Publishing Co., Singapure, 1989.

\bibitem{Fi25}
R.~A.~Fisher,
{\it Proc.~Philos.~Camb.~Soc.} {\bf 22} (1925) 700



\bibitem{Fo28}
V.~Fock, {\it Z.~Phys.} {\bf 49} (1928) 323

\bibitem{Ka48}
S.~Kakutani,
{\it Ann.~of Math.} {\bf 49} (1948) 214



\bibitem{Sa65}
S.~Sakai,
{\it Bull.~Am.~Math.~Soc.} {\bf 71} (1965) 149

\bibitem{Bu69}
D.~J.~C.~Bures,
{\it Trans.~Amer.~Math.~Soc.} {\bf 135} (1969) 199




\bibitem{Do74}
W.~Donoghue, Monotone matrix functions and analytic continuation.
Berlin, Heidelberg, New York: Springer 1974







\bibitem{KA80}
F.~Kubo, T.~Ando,
{\it Math.~Ann.} {\bf 246} (1980) 205


\bibitem{book82.5}
A.~S.~Holevo,
{\it Probabilistic and Statistical Aspects of Quantum Theory},
North-Holland, Amsterdam, 1982







\bibitem{IJKK82}
R.~S.~Ingarden, H.~Janyszek, A.~Kossakowski, and T.~Kawaguchi,
{\it Tensor, N.~S.} {\bf 37} (1982) 105 




\bibitem{Ko83}
H.~Kosaki,
{\it Proc.~Am.~Math.~Soc.} {\bf 89} (1983) 285

\bibitem{Si83}
B.~Simon,
{\it Phys.~Rev.~Lett.} {\bf 51} (1983) 2167

\bibitem{Be84}
M.~V.~Berry,
{\it Proc.~Royal.~Soc.~Lond.} {\bf A 392} (1984) 45

\bibitem{WZ84}
F.~Wilczek and A.~Zee,
{\it Phys.~Rev.~Lett.} {\bf 52} (1984) 2111




\bibitem{Uh86a}
A.~Uhlmann,
{\it Rep.~Math.~Phys.} {\bf 24} (1986) 229


\bibitem{AA87}
Y.~Aharonow  and J.~Anandan,
{\it Phys.~Rev.~Lett.} {\bf 58} (1987) 1593


\bibitem{AS87}
J.~Anandan and L.~Stodolsky,
Phys.~Rev. D35 (1987) 2597




\bibitem{SB88}
J.~Samuel and R.~Bhandari,
{\it Phys.~Rev.~Lett.} {\bf 60} (1988) 2339;



\bibitem{Uh89b}
A.~Uhlmann,
Ann.~der Physik, 46 (1989) 63 - 69


\bibitem{DG90}
L.~Dabrowski, H.~Grosse,
{\it Lett.~Math.~Phys.} {\bf 19} (1990) 205





\bibitem{Uh91a}
A.~Uhlmann,
{\it Lett.~Math.~Phys.} {\bf 21} (1991) 229




\bibitem{Uh92a}
A.~Uhlmann,
Phys.~Lett. A 161 (1992) 329 - 331

\bibitem{DR92a}
J.~Dittmann and G.~Rudolph,
{\it J.~Math.~Phys.} {\bf 33} (1992) 4148 

\bibitem{DR92b}
J.~Dittmann and G.~Rudolph,
{\it J.~Geom.~and Phys.} {\bf 10} (1992) 93


\bibitem{Hu92b}
M.~H\"ubner,
{\it Phys.~Lett.} {\bf A 163} (1992) 239

\bibitem{Uh92b}
A.~Uhlmann,
in: Quantum Groups and Related Topics, {\it eds.~R.~Gielerak et al.},
Kluwer Acad.~Publishers, 1992, 267

\bibitem{OP93}
M.~Ohya, D.~Petz,
{\it Quantum Entropy and Its Use}.
Texts and Monographs in Physics,
Berlin: Springer-Verlag, 1993.

\bibitem{Di93a}
J.~Dittmann,
{\it Seminar Sophus Lie} {\bf 3} (1993) 73 

\bibitem{Hu93a}
M.~H\"ubner,
{\it Phys.~Lett.} {\bf A 179} (1993) 226 


\bibitem{Ha93}
H.~Hasegawa,
{\it Rep.~Math.~Phys.} {\bf 33} (1993) 87




\bibitem{BC94}
S.~L.~Braunstein, C.~M.~Caves,
{\it Phys.~Rev.~Lett.} {\bf 72} (1994) 3439



\bibitem{FC95}
C.~A.~Fuchs, C.~M.~Caves,
{\it Open Sys.~Inf.~Dyn.} {\bf 3} (1995) 1

\bibitem{Pe95}
D.~Petz,
Information-Geometry of Quantum States,
Budapest, preprint 39/1995

\bibitem{PH95a}
D.~Petz, C.~Hasegawa,
{\it Lett.~Math.~Phys.} {\bf 38} (1996) 221


\bibitem{Pe96}
D.~Petz,
{\it Linear Alg.~Appl.} {\bf 244} (1996) 81

\bibitem{PS96}
D.~Petz and C.~Sudar,
{\it J.~Math.~Phys.} {\bf 37} (1996) 2662






\bibitem{BR97}
R.~Bhatia and P.~Rosenthal
{\it Bull.~London Math.~Soc.} {\bf 29} (1997) 1

\bibitem{To93}
G.~Rudolph and T.~Tok,
Rep.~Math. Phys.~ {\bf 39}
(1997), 433-446

\bibitem{Di98}
J.~Dittmann,
quant-ph/9806018

\end{thebibliography}
\end{document}